# Integrated bolometric photodetectors based on transparent conductive oxides for near- to mid-infrared wavelengths

J. Gosciniak[1*]

[1]ENSEMBLE3 sp. z o.o., Wolczynska 133, 01-919 Warsaw, Poland
Corresponding authors: *jeckug10@yahoo.com.sg

**Abstract**
On-chip photodetectors are essential components in optical communications as they convert light into an electrical signal. Photobolometers are type of photodetector that functions through a resistance change caused by electronic temperature fluctuations upon light absorption. They are widely used in the broad wavelength range from UV to MIR and can operate on a wide material platform. In this work, I introduce a novel waveguide-integrated bolometer that operates in a wide wavelength range from NIR to MIR on the standard material platform with the transparent conductive oxides (TCOs) as the active material. This material platform enables the construction of both modulators and photodetectors using the same material, which is fully CMOS compatible and easily integrated with passive on-chip components. The photobolometers proposed here consist of a thin TCO layer placed inside the rib photonic waveguide to enhance light absorption and then heat the electrons in the TCO to temperatures above 1000 K. This rise in electron temperature leads to decreasing electron mobility and consequential electrical resistance change. In consequence, a responsivity exceeding 10 A/W can be attained with a mere few microwatts of optical input power. Calculations suggest that further improvements can be expected with lower doping of the TCO, thus opening new doors in on-chip photodetectors.

**Introduction**
Photonic Integrated Circuits (PICs) enable the large-scale integration of many different components, ranging from modulators, filters, splitters, (de)multiplexers to photodetectors, on a single chip to provide a more energy-efficient way to increase the speed and capacity of data networks and reduce costs [1-3]. Among different components, photodetectors and modulators are the main building blocks of the optoelectronic link that provides signal conversion between electronic and photonic domains [4, 5], therefore the monolithic integration of electro-optical components, i.e., modulators and photodetectors, is highly required to increase the chip functionality. Compared to modulators, which perform electrical-to-optical signal conversion, photodetectors perform inverse conversion, i.e., from the optical to the electrical domain. In addition, integrating the detector on-chip can improve responsivity by reducing the volume of active material that can generate thermal noise [6].
So far, most of the efforts have been focused on the near-infrared (NIR) PICs in the context of data communication, so the range of possible photodetector arrangements is very wide [7]. In contrast, mid-infrared (MIR) PICs are still in the early stages of development, lagging far behind their NIR counterparts [7, 8]. Thus, the field of material platform for MIR photodetectors is very limited to a few semiconductors such as PbTe, black phosphorus, tellurene and graphene [9]. However, a huge growth of MIR PICs is expected in the coming years, mainly due to the development of sensors for gas detection, biological systems, security, and industrial applications, [https://mirphab.eu].
Photonic integrated circuits in the MIR require new devices that can operate in the MIR wavelength range and are therefore likely to be based on a new material platform [8]. One such device is a photodetector, which converts an optical signal into an electrical signal and is an essential component in on-chip O-E conversion. However, it must meet several important requirements, such as compatibility with CMOS technology, operation over a wide wavelength range, and no need for cooling, which adds complexity and cost to the system [6]. In comparison, most of the previously proposed photodetectors in the MIR wavelength range are either expensive to manufacture, do not operate over a wide wavelength range, or are impractical because they need to be cooled to low temperatures. Thus, the search for the MIR photodetector is still ongoing. The solution may be thermal detectors that convert heat into electricity [10-14]. They require an absorbing material that absorbs light to generate hot carriers and then converts them to electricity.

Transparent conductive oxides (TCOs), which belong to the epsilon-near-zero (ENZ) materials, seem to be an excellent material for such a task, as they can absorb energy under a wide range of wavelengths [15-20]. They possess large permittivity tunability under an applied voltage or light illumination [21-26], allowing the material properties to be tuned to specific requirements. They also feature low optical loss, fast switching time and low switching voltage. In addition, they are CMOS compatible and can be mass-produced using standard fabrication methods [15, 28, 29, 31]. Thus, they meet all the technological requirements. In addition, TCOs have already been shown to offer many advantages, such as being an active material in electro-absorption modulators capable of achieving even unit-order index modulation and showing strong modulation at ENZ wavelengths [23-25].

**Main text**
**Photodetectors – from NIR to MIR**
Progress in the near infrared (NIR) wavelength range has been largely dictated by silicon and its transparency window in the NIR-IR [27]. Silicon is an indirect bandgap semiconductor with a bandgap energy of 1.1 eV, and therefore exhibits poor emission and detection properties in the NIR, which enables applications for telecommunications due to low propagation losses. After many years of extensive development, silicon technology benefits from reliable and high-volume manufacturing and offers low cost, high performance and compact circuits. In contrast to NIR, MIR systems need to move away from standard silicon-based materials and rely on an alternative material platform that can be integrated on silicon [8]. As a result, MIR technology is not as mature as its NIR counterpart, making production costs significantly higher. However, the manipulation of light at mid-infrared (MIR) wavelengths is essential for many applications ranging from biological sensing, heat transfer, photothermal conversion to free-space optical communications [6, 9].

The need for MIR photodetectors follows directly from the second law of thermodynamics, which states that the overall entropy of the system increases. As a consequence, any biological or artificial system, in order to function, must absorb "high quality" energy in the form of light, food, or well-organized energy sources, process it, and finally dissipate "low quality" energy, i.e. energy that is reduced to a state of maximum disorder, usually in the form of thermal radiation, i.e. heat, ranging from 2 µm to 100 µm, or gases. Thus, the second law of thermodynamics states that the quality of energy is irreversibly degraded and removed from the system as heat or gas. As a result, all devices operating in the MIR wavelength range are essential for monitoring all issues related to our life.

**Transparent conductive oxides platform**
Here, we propose a unique material platform for the realization of NIR and MIR photodetectors based on transparent conductive oxides (TCOs). The family of transparent conductive oxides (TCOs) is broad, ranging from indium tin oxide (ITO), indium zinc oxide (IZO), cadmium oxide (CdO), gallium zinc oxide (GZO), aluminum zinc oxide (AZO) to mention only some of the most popular compounds [28, 29]. Thus, they can operate in a wide wavelength range from UV to MIR. Furthermore, the operating wavelength can be tuned either during a fabrication process under appropriate doping of oxide compounds or under electrical and/or optical signals, which opens new possibilities in terms of optoelectronic devices with tunable characteristics [28].

In this work, I have focused on three TCO materials: ITO, GZO and AZO, which can serve as an example. As for the zinc oxide (ZnO) compounds presented in this paper, they were highly doped with 6% Ga and 10% Al, respectively; thus, the ENZ wavelengths were shifted to shorter wavelengths. It is well known that undoped ZnO exhibits the ENZ point at the wavelength of 14-20 µm [29], therefore the ENZ wavelength can be tuned under doping from the long-wave infrared (LWIR), the mid-wave infrared (MWIR) to the short-wave infrared (SWIR) wavelength range. In addition, ZnO is a very interesting compound as it is biocompatible, biodegradable and biosecure and is already considered for sensing, optoelectronic and storage applications [30-32]. Furthermore, it is able to efficiently convert heat into electricity, thus serving as a thermoelectric device [10, 11].

**Waveguide arrangement**

By integrating the TCO material with a silicon-on-insulator (SOI) waveguide, it is possible to greatly enhance the TCO absorption and thus the corresponding photodetection efficiency. In the proposed waveguide-integrated device, the mode from a photonic waveguide efficiently couples to the TCO layer via the evanescent field, leading to optical absorption and the generation of hot carriers, which are then detected by two metal electrodes located on opposite sides of the waveguide at a distance as small as 1 µm (Fig. 1a).

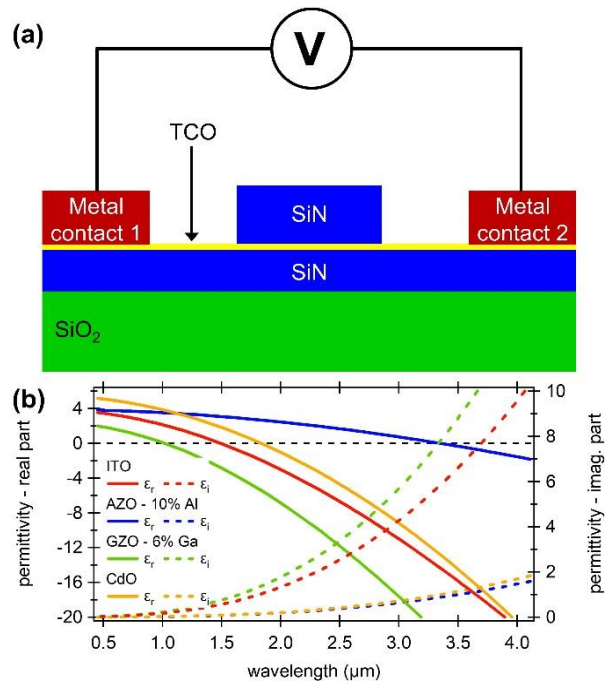

**Fig. 1**. (a) Geometry of the proposed TCO-based photodetector organized in the photonic rib waveguide arrangement. (b) Dispersion of complex permittivity (real and imaginary parts) of ITO, GZO, AZO and In:CdO as a function of wavelength.

Figure 1b shows the complex permittivity (real and imaginary parts) as a function of wavelength for four TCO materials - ITO, CdO, AZO and GZO. As shown earlier, the materials with higher plasma frequency are characterized by lower ENZ wavelength due to higher carrier concentration. For example, the ENZ wavelength for GZO with 6% Ga is $\lambda=1$ µm, while for ITO and CdO it is $\lambda=1.5$ µm and $\lambda=1.8$ µm, respectively. The ENZ wavelength for ZnO doped with 10% Al is the highest and is located at the wavelength $\lambda= 3.3$ µm. However, decreasing the dopant concentration will shift the ENZ point to longer wavelengths, as the ZnO compound without doping (Al or Ga) shows the EZN wavelength at around $\lambda=14-20$ µm [33].

**Optical properties of TCO materials**

The complex permittivity of TCO materials allowed us to calculate the mode effective index and mode power attenuation for different waveguide configurations as shown in Figure 1a. The real and imaginary parts of the permittivity for TCO materials were calculated using the Drude formula:

$$\varepsilon = \varepsilon_\infty - \frac{\omega_p^2}{\omega^2 - i\omega\Gamma} \quad (1)$$

where $\varepsilon_\infty$ is the permittivity due to bound electrons, $\omega_p$ is the plasma frequency, $\omega$ is the angular frequency of light, and $\Gamma$ is the scattering rate, damping factor. The plasma frequency can be expresses as:

$$\omega_p^2 = \frac{N_c e^2}{\varepsilon_0 m^*} \quad (2)$$

where $N_c$ is the carrier concentration, $e$ is the electric charge, $\varepsilon_0$ is the background permittivity, and $m^*$ is the effective mass. Thus, it depends directly on both the carrier concentration and the effective mass of the electrons in the conduction band of TCO. The properties of the electrons in the conduction band are determined by the conduction-electron energy band. Since TCO materials are characterized by the non-parabolic conduction band, the energy-wave vector (E-k) dispersion relation of the conduction band is expressed by:

$$\frac{\hbar^2 k^2}{2m^*} = E + CE^2 \quad (3)$$

where $\hbar$ is the reduced Planck constant, $k$ is the electron wavevector, $m^*$ is the electron effective mass at the conduction band minimum, $E$ is the electron energy at the conduction band minimum, and $C$=0.4191 eV$^{-1}$ is the constant known as the first-order nonparabolicity factor ($C$=0 for the parabolic band) [34-36]. Nonparabolicity is defined here by an additional term $CE^2$ in the E-k diagram, which introduces a nonlinearity and causes the effective mass $m^*$ to become a function of the electron density $N_c$, i.e. the effective mass increases with increasing carrier concentration under interband absorption of light or electrical doping [34]. It is described by the density of states:

$$\rho_e(E) = \frac{1 + 2CE}{2\pi^2} \left(\frac{2m^*}{\hbar^2}\right)^{3/2} \sqrt{E(1 + CE)} \quad (4)$$

As the carrier density increases, the conduction band fills up, causing the Fermi energy to increase and involving higher band regions in determining the optical properties of the TCO, which are characterized by higher effective mass [34]. Thus, interband absorption of light causes the carrier concentration to increase, which increases the plasma frequency and blueshifts the ENZ wavelength. However, for materials with high nonparabolicity, in some special cases, the increases in effective mass can dominate the carrier concentration increase and the interband transition can cause an overall redshift of the ENZ wavelength, inverting the nonlinearity [37]. Consequently, the overall magnitude and direction of the plasma frequency shift and the nonlinearity change depend on the nonparabolicity of the conduction band and the carrier density.

Furthermore, the effective mass can be tuned by intraband absorption of light. When light is absorbed, the energy is transferred to the electrons in the TCO, resulting in an increase in the electron temperature $T_e$. As a result, the electrons move to higher energy states in the conduction band with higher effective mass $m^*$ [36]. At the same time, the plasma frequency $\omega_p$ decreases. As the TCO materials are characterized by lower electron density compared to noble metals, their electron heat capacity is much smaller [22, 35, 36], therefore the electrons in TCOs heat up more and cool down faster compared to noble metals. For the same amount of energy provided to the system, the electrons in the TCO material with the lower carrier concentration heat up to higher temperatures, resulting in a higher effective mass. In the absence of light, the relaxation process results in the transfer of energy from the electrons to the lattice in the form of heat.

It is important to note that for a low light frequency and low carrier concentration, the scattering contribution can play a significant role as it depends on both the effective mass and the carrier mobility and is expressed by:

$$\Gamma = \frac{e}{m^* \mu_{mob}} \tag{5}$$

where $\mu_{mob}$ is the carrier mobility. A longer ENZ wavelength, as in the case of AZO (Fig. 1b), corresponds to a lower free carrier density and therefore a lower Fermi level. It should be remembered that for the same amount of absorbed energy, electrons at a lower Fermi level undergo a larger change in effective mass than electrons at a higher Fermi level, resulting in a larger nonlinearity [37]. Furthermore, as the effective mass increases, the carrier mobility decreases [38]. However, this relationship is not linear and depends on many contributions, such as the carrier concentration in the TCO and therefore the Fermi energy level, the quality of the fabricated TCO thin films, thus the scattering rate can increase or decrease under different working conditions. It has been experimentally confirmed that an increase in the effective mass under intraband absorption of light causes the plasma frequency to decrease and, at the same time, the scattering rate to increase [38. 39]. Since both the plasma frequency (Eq. 2) and the scattering rate (Eq. 5) are inversely proportional to the effective mass, the increase of the scattering rate under the increase of the light absorption results in the decrease of the mobility.

Longer ENZ wavelength, as in the case of AZO, corresponds to lower free carrier density and therefore lower Fermi level. As mentioned above, the electrons at lower Fermi level undergo a larger change in effective mass compared to electrons at higher Fermi level for the same amount of absorbed power, which leads to a larger nonlinearity [37].

**Field enhancement in thin TCOs**

The change in effective mass under intraband absorption of light is directly dependent on the electron temperature and thus on the energy provided to the system, i.e., the TCO material. The amount of light absorbed by the TCO thin film depends on the coupling arrangement, since a thin film of TCO material exhibits optical anisotropy and polarization-dependent resonant behavior at ENZ wavelengths. Depending on the polarization of the light, i.e., the electric field polarized along the width (TE mode) or height (TM mode) of the TCO layer, a different electric field strength can be achieved in the TCO. Consequently, for the TM-polarized light, a high electric field enhancement, and thus energy, is expected because the displacement $D=\varepsilon E$ is continuous across the interface, leading to the huge electric field enhancement in the ENZ materials, which is much higher than in the conventional dielectrics. The high field enhancement in thin TCO films near the ENZ point results directly from the boundary condition $\varepsilon_0 E_{0\perp} = d\varepsilon_1 E_{1\perp}$, which requires the phase matching of the total electric fields on both sides of the interface. Here, $d$ is the thickness of the TCO film, $\varepsilon_0$ and $\varepsilon_1$ are the permittivities of air and TCO film, respectively, and $E_{0\perp}$ and $E_{1\perp}$ are the normal components of the electric field in the waveguide and TCO, respectively. Thus, for a small $\varepsilon_1$ near the ENZ region, the field normal to the TCO film is strongly enhanced. Furthermore, as the thickness of the TCO is reduced and more of the optical field is concentrated within the TCO layer, the absorption increases. As a result, more energy is provided to the TCO, which further enhances the electric field in the TCO and causes an increase in the electron temperature in the conduction band. This in turn leads to an increase in effective mass and a corresponding increase in permittivity, moving it closer to the ENZ point. As a result, extremely high absorption can be achieved in the TCO near the ENZ wavelength (Fig. 2), which can lead to an increase in photoresponsivity [26].

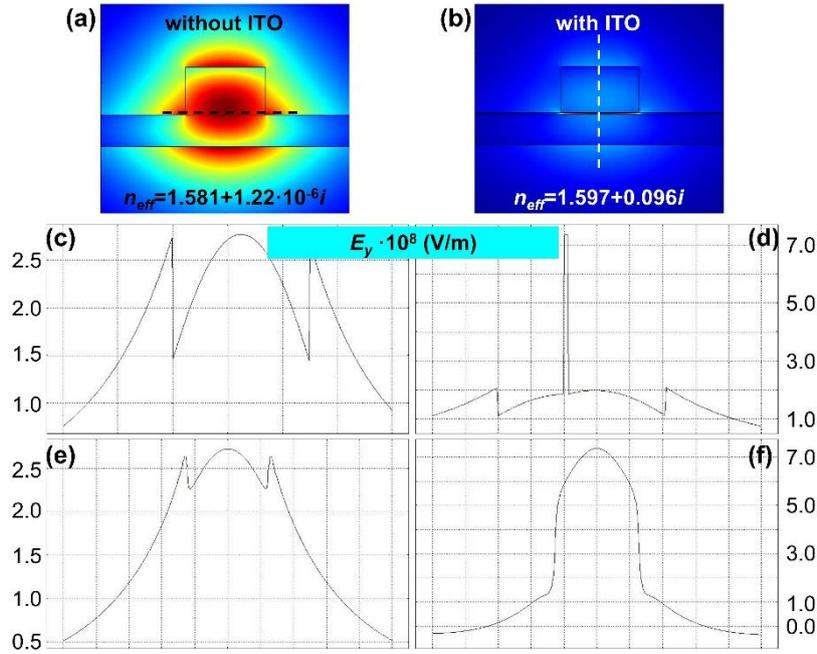

**Fig. 2**. (a, b) Electric field profile ($E_y$ component) for structure (a) without ITO layer and (b) with the ITO layer at the ENZ wavelength. (c, d) Electric field profile (Ey component) in normal to the TCO layer (dashed white line in b) and (e, f) along the TCO layer (black dashed line in a) for structure (c, e) without TCO layer and (d, f) with TCO layer.

**TCO in a photonic waveguide arrangement**

For all waveguide arrangements investigated in this paper, the TCO thickness was kept constant at 10 nm. For a shorter wavelength of 1550 nm, the TCO was placed in the SiN rib waveguide arrangement with SiN rib dimensions of $w$=500 nm and $h$=300 nm, while the SiN rib thickness was kept at $t$=200 nm. To ensure good guiding properties, the dimensions of the SiN rib waveguide were increased to $w$=800 nm, $h$=450 nm, and $t$=350 nm for a longer wavelength of 2500 nm. However, when SiN was replaced by Si, the dimensions were reduced to $w$=400 nm, $h$=300 nm, and $t$=200 nm. It should be noted that simulations were performed for the cases where the external electrodes were placed on both sides of the ridge at a distance of 1 μm from both sides of the ridge as shown in Fig. 1. No perturbation of the propagation modes was observed in all these cases.

Permittivity calculations for AZO showed that the ENZ point shifts to lower carrier concentrations as the excitation wavelength increases, as shown in Fig. 3a. To compensate for the increase in wavelength, i.e. the decrease in light frequency, the plasma frequency must decrease, which can be achieved by decreasing the carrier concentration. Furthermore, a longer ENZ wavelength, as in the case of AZO (Fig. 1b), corresponds to a lower free carrier density and therefore a lower Fermi level. The electrons at the lower Fermi level undergo a larger change in effective mass compared to electrons at the higher Fermi level for the same amount of absorbed energy, resulting in a larger nonlinearity [37].

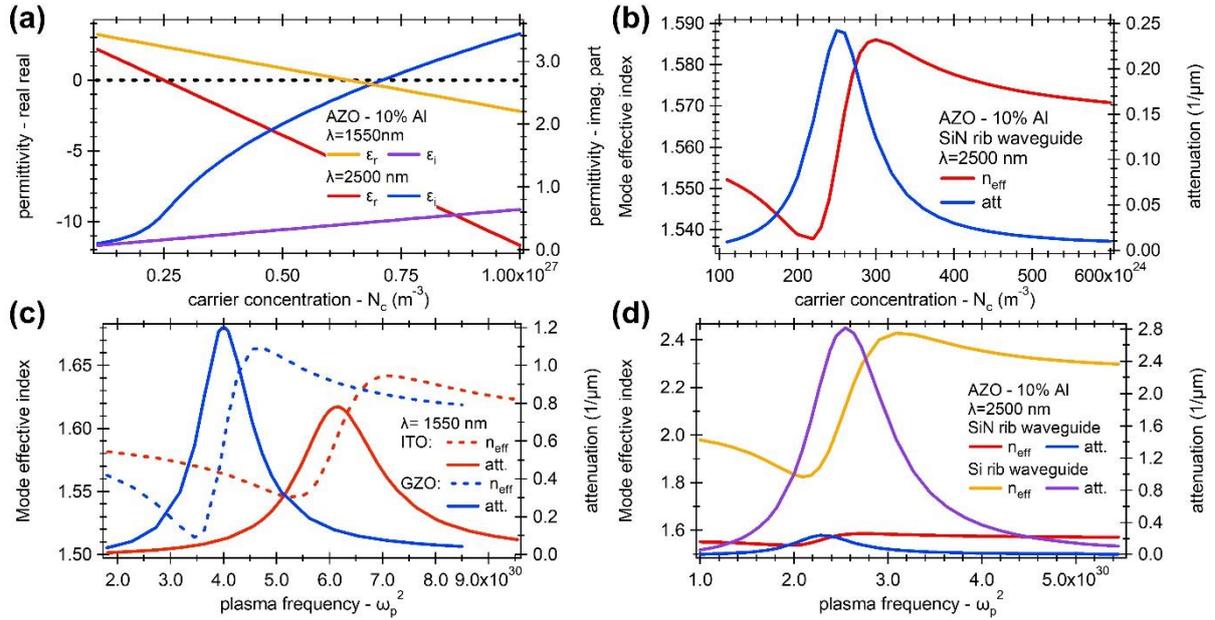

**Fig. 3**. (a) Permittivity (real and imaginary) of the AZO film as a function of carrier concentration and two different wavelengths of light – λ=1550 nm and λ=2500 nm. (b) Mode effective index and mode attenuation as a function of carrier concentration for wavelength of 2500 nm. (c, f) Mode effective index and mode power attenuation as a function of plasma frequency for (c) ITO and GZO and (d) AZO.

As observed in Fig. 3d for AZO placed in the Si rig waveguide arrangement, a more confined mode provides a higher electric field enhancement in the TCO. As more energy is provided to the TCO, the electron-electron and electron-photon scattering time decreases due to the higher probability of collisions between photoexcited electrons and the overall scattering rate increases. Thus, for the same carrier concentration in the TCO, the scattering contribution in the low-frequency permittivity part of the Drude formula decreases, so the ENZ point shifts to higher carrier concentration to compensate for an increase in the scattering rate at higher power coupled to the TCO. This shift can be significant, as observed in Fig. 3d, and can be equivalent to increasing the carrier concentration by $0.3 \cdot 10^{26}$ m$^{-3}$. For a highly confined mode operating at a wavelength of 2500 nm, as in the case of the Si rib waveguide, the electric field enhancement exceeds $|E_y|=3 \cdot 10^9$ V/m, while for a less confined mode with the SiN rib waveguide it is an order of magnitude lower and reaches about $|E_y|=1.6 \cdot 10^8$ V/m.

**Electron system dynamics in the TCO**

The intraband absorption of light at a given frequency results in the excitation of electrons from an initial energy level $E_i$ to the final energy level $E_f=E_i+\hbar\omega$. The level of electron energy increase depends on the energy of the absorbed pulse, which is proportional to the local energy density or $|E(t)|^2$, i.e. the local electric field vector $E(t)$ in the TCO material, as shown in Fig. 2. Almost immediately after the light is turned off, the electrons start to relax back to the initial state under different relaxation processes. First, the electrons are thermalized by the electron-electron collisions on a time scale of tens of femtoseconds, leading to a convergence of the non-thermal population to the thermalized Fermi-Dirac distribution. In the next step, the electrons relax by collisions with phonons on a time scale of hundreds of femtoseconds, which is much slower compared to the electron-electron relaxation time. It results from a large mismatch between the Fermi momentum and the Debye momentum in TCOs.

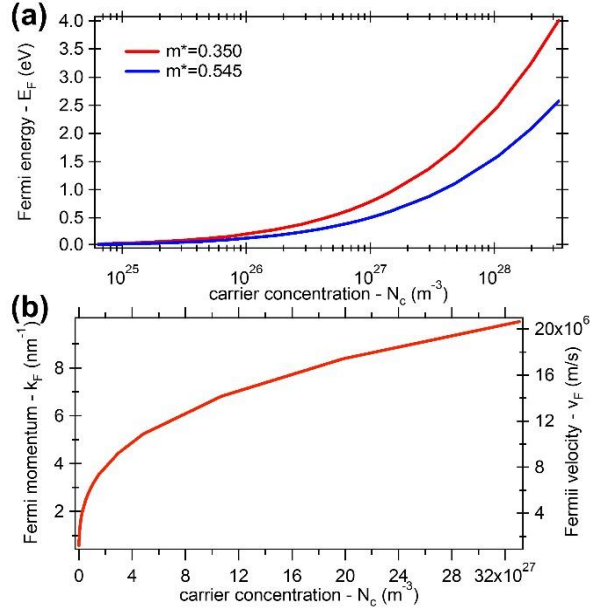

**Fig. 4**. (a) Fermi energy and (b) Fermi momentum and Fermi velocity distribution at different carrier concentrations in TCO materials.

For example, for the electron concentration $N_c=4\cdot10^{26}$ m$^{-3}$ in ITO (effective mass $m^*=0.35$), the Fermi energy was calculated to be $E_F=0.47$ eV (Fig. 4a). It corresponds to a Fermi momentum $k_F\approx2.28$ nm$^{-1}$ (Fig. 4b), which is much smaller than the Debye momentum of TCO, calculated to be $k_D\approx19.0$ nm$^{-1}$ for the Debye temperature $T_D=900$ K. Thus, due to the momentum conversion, a large number of phonons are forbidden (not allowed) to interact with the electrons. As a consequence, the electron-phonon relaxation time increases. In comparison, the carrier concentration in noble metals is much higher ($5.9\cdot10^{28}$ m$^{-3}$ for gold), therefore the thermalization of electrons, i.e. electron-electron collisions, is about 10 times slower compared to TCO materials [35, 36]. This results in a much higher temperature rise in TCOs compared to noble metals. However, the electron-phonon collision rate in noble metals is similar to that in TCO materials, despite the much higher Fermi energy, $E_F=15.8$ eV, mainly due to the higher Fermi momentum ($k_F\approx12$ nm$^{-1}$ in gold) compared to the Debye momentum ($k_D\approx6.85$ nm$^{-1}$). Thus, almost all electrons in noble metals can interact with all phonons, resulting in a shorter electron-phonon scattering time. Furthermore, a similar level of electron-phonon collision rates in TCOs and noble metals translates into a similar electron-phonon coupling coefficient $G_{e-ph}$, which was calculated to be $G_{e-ph}=3\cdot10^{16}$ Jm$^{-3}$K$^{-1}$s$^{-1}$ for ITO and $G_{e-ph}=2.5\cdot10^{16}$ Jm$^{-3}$K$^{-1}$s$^{-1}$ for gold [36].

The electron temperature rise in TCO is determined by the electronic heat capacity $C_e\approx\gamma_e T_e$, where $T_e$ is the electron temperature in TCO under light illumination and $\gamma_e$ is the Sommerfeld coefficient. The $\gamma_e$ for TCO ($\gamma_e=12.7$ Jm$^{-3}$K$^{-2}$ for ITO) is smaller than for noble metals ($\gamma_e=67.6$ Jm$^{-3}$K$^{-2}$ for gold) [35], which is related to the lower electron density. Thus, the electron heat capacity of ITO is at least 5 times smaller than that of gold. As a result, for the same power provided to the system, $E_{abs}$, and ignoring the electron-phonon heat transfer, the electron temperature scales with $(E_{abs}/\gamma_e)^{1/2}$, and was calculated to be $T_e=2100$ K and $T_e=900$ K for ITO and gold, respectively [35].

Besides electron-electron and electron-phonon scattering, the third process that contributes to the relaxation of electrons and thus to the dielectric constant damping terms should be considered, i.e. electron-impurity collisions. However, unlike the other processes, the electron-impurity relaxation rate is independent of the electron distribution [35]. Finally, the complex damping term $\Gamma$ in the permittivity formula of TCO materials (Eq. 1 and 5) consists of three relaxation processes and depends on the energy provided to the material. It has been experimentally confirmed that the damping term $\Gamma$ increases linearly with the electron temperature $T_e$ [22]. The energy provided to the electrons leads to an increase of the electron temperature $T_e$ and thus to an increase of the real part of the permittivity $\varepsilon'$. Therefore, the ENZ resonance is shifted from the current frequency so that the field and the absorptivity decrease. Thus, the dependence of the total absorbed energy and hence the maximum electron and photon temperature is linear only in a very narrow intensity range [36].

For short light pulses, most of the cooling occurs through the phonons, so almost all of the absorbed energy is transferred to the phonon subsystem before escaping to the environment [35]. Consequently, as the electrons cool down, the phonon temperature increases. In contrast, for light pulses longer than the electron-electron relaxation (thermalization) time $\tau_{e-e}$, the electron temperature in TCO behaves dynamically. First, the electron temperature decay rate is related to the electron heat capacity $C_e$, the electron-photon coupling coefficient Ge-ph, and the temperature difference between electrons and phonons [36]. As shown previously, the electron temperature decay rate is faster at higher electron temperatures and is proportional to the temperature difference between electrons and phonons [35, 36].

**Speed limitations in TCO-based photodetectors**

As it was mentioned above, the time of decrease of the electron temperature in the TCO is determined by the ratio:

$$\tau = \frac{C_e}{G_{e-ph}(T_e - T_{ph})} \tag{6}$$

where $C_e$ is the electron heat capacity, $G_{e-ph}$ is the electron-phonon energy coupling coefficient, $T_e$ is the electron temperature and $T_{ph}$ is the phonon temperature. The decay time of electron temperature in the TCO corresponds to the thermal response time of the TCO bolometer. For example, for ITO the $G_{e-ph}$ was calculated at $G_{e-ph}$=2.5·10$^{16}$ Jm$^{-3}$K$^{-1}$s$^{-1}$, while heat capacity for electron and phonon at $C_e$=1.0·10$^4$ Jm$^{-3}$K$^{-1}$ (at $T_e$=600 K) and $C_{ph}$=2.54·10$^6$ Jm$^{-3}$K$^{-1}$ for $T_e$-$T_{ph}$=300K, respectively [**35**]. Thus, the thermal response time was estimated at $\tau$=102 ps what corresponds to a 3 dB cutoff frequency $f_{3dB}$=9.8 GHz and is dominated by a phonon heat capacity. A lowering of the electron density in the TCO can bring some additional benefits. It reduces the electron heat capacity therefore the electrons can heat much more and cool down faster. Furthermore, as the electron heat capacity is reduced, the response time decreases [**35, 36**].

**Electron temperature in TCO vs optical power coupled to the photodetector**

To calculate the performance of the photobolometric photodetector the electron temperature increase $\Delta T_e$ in TCO under the absorption of light should be estimated. The generated photocurrent in a bolometric photodetector is determined by the electron temperature change $\Delta T_e$ in the TCO material under a light coupled to a photodetector according to the equation:

$$I_{ph} \approx \frac{W \cdot h}{L}\left(\left.\frac{\partial \sigma}{\partial T_e}\right|_{T_e=T_0} \Delta T_e\right) U_b \tag{7}$$

where $L$ is the length of the photodetector, $W$ and $h$ are the width and height of the photodetector, $\partial\sigma/\partial T_e$ is the rate of conductivity change, defined by $\sigma=ne\mu_{mob}$, under a change of electron temperature. The photodetector proposed in this paper operates under an intraband absorption of light, i.e. for wavelengths in the NIR and MIR range, therefore only the electric mobility can be affected by light and no change in carrier concentration is expected. Consequently, the change in conductivity is related to the rate of change in electrical mobility. Therefore, it is necessary to evaluate the electron temperature rise in the TCO material under absorbed light and its influence on the electrical mobility.

The temperature of the electrons in the TCO material as a function of the absorbed power is governed by the heat transfer equation:

$$C_e \frac{dT_e}{dt} T_e + G_{e-ph}(T_e - T_{ph}) - \frac{P}{A} = 0 \qquad (8)$$

where $T_e$ is the electron temperature, $P$ is the absorbed power by TCO, and $A$ is the unit area of TCO that absorbs light. The analytical solution to the heat equation along a photodetector length $L$ is:

$$\Delta T_e(y) = T_e(y) - T_{ph} = \frac{\sinh((0.5L - |y|)\xi)}{2\cosh(0.5L \cdot \xi)} \frac{P \cdot \tau}{A} \frac{\xi}{C_e} \qquad (9)$$

where $\xi=(G_{e-ph}/\kappa)^{0.5}$, while the inverse of $\xi$, $1/\xi=1/(G_{e-ph}/\kappa)^{0.5}$ is the cooling length of the electrons in TCO, $\kappa$ is the electron thermal conductivity of TCO, $\tau$ is the time of the light pulse, $A$ is the area of the TCO layer under illumination, $P$ is the power coupled to the TCO layer, $C_e$ is the electron heat capacity of TCO. Assuming the thermal conductivity of the ITO material at $\kappa=0.5$ Wm$^{-1}$K$^{-1}$ and the input power of only 20 µW, i.e., the power coupled to the photodetector, the mode power attenuation close to the to the ENZ point calculated at $att=0.8$ µm$^{-1}$ (Fig. 3a), and the mode propagation distance $L=2$ µm, the maximum electron temperature was calculated at $T_e=7450$ K. In comparison, for a thermal conductivity $\kappa=20$ Wm$^{-1}$K$^{-1}$ with other parameters kept as before, the maximum electron temperature was much smaller and was calculated at $T_e=1180$ K. The propagation distance of 2 µm is sufficient to ensure that almost all of the power provided to a photodetector is absorbed by the TCO layer leading to the electron temperature increase in the TCO. In all these cases, the pulse duration was kept at $\tau=50$ ps. The pulse duration and power define the energy provided to the system, while the width and height of the TCO layer define the energy density. Thus, to increase the energy density provided to the system under the same power of light, either the pulse duration should be increased or the dimensions of the TCO layer should be minimized. As shown previously [36], for the same amount of power coupled to the TCO, the shorter pulse provides a higher efficiency of electron heating because less energy is transferred from the electron to the phonon subsystem. As a result, the electron temperature rises rapidly and much faster than in the case of the longer pulse.

However, the longer pulse provides more energy, so the total electron temperature is higher. In consequence, it leads to a larger change in the real part of TCO permittivity, $\varepsilon'$, and a less pronounced drop in the dynamics of the imaginary part of the TCO permittivity, $\varepsilon''$ [36].

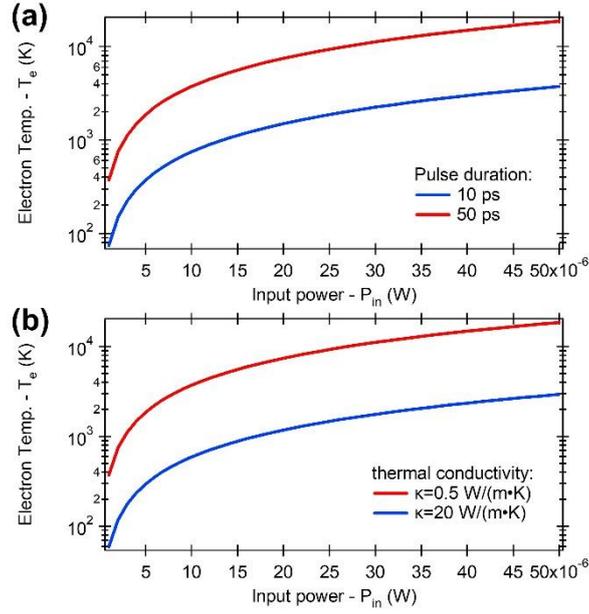

**Fig. 5**. (a, b) Electron temperature $T_e$ as a function of input power for (a) different pulse durations of 10 ps and 50 ps and (b) thermal conductivity of TCO of 0.5 W/(m·K) and 20 W/(m·K).

Here, the calculations were performed for ITO with a carrier concentration of $N_c$=1.5·10$^{27}$ m$^{-3}$, which corresponds to a Fermi energy of $E_F$=0.88 eV. However, as shown above, the lower carrier concentration results in a lower Fermi energy and lower Fermi momentum (Fig. 4), thus increasing the mismatch between the Fermi momentum and the Debye momentum which leads to an increased number of phonons that are prohibited from interacting with the electrons. In consequence, the electron-phonon relaxation time increases, resulting in a higher electron temperature. As a result, further improvement is possible with a proper choice of carrier concentration in the TCO materials.

**Electron carrier mobility and electron effective mass**

To be consistent with other papers [35, 36, 38], the calculations were performed for ITO thin film with a high carrier concentration of 1.5·10$^{27}$ m$^{-3}$, therefore, the initial effective mass was calculated to be $m^*$=0.3964. The experimental data reported in Ref. [38] for ITO at the same carrier concentration of 1.5·10$^{27}$ m$^{-3}$ showed that the electron mobility decreases with increasing the electron temperature $T_e$, with the curve fit described by equation:

$$\mu_{mob} = 18.3 - 2.13 \cdot 10^{-5} T_e^{1.53} \tag{10}$$

As pointed out by the authors [38], the exponential index 1.53 at $T_e$ is close to 3/2 indicating the ionized impurity scattering as the dominant scattering mechanism [15]. The equation shows that at the room temperature the electron mobility was measured to be $\mu_{mob}$=18.3 cm$^2$/(V·s), which is in good agreement with the other measurements [34, 40]. Similar electron mobility values have been measured for other TCO materials. For example, the measurements on AZO thin film at room temperature showed a mobility of 16.4 cm$^2$/(V·s) at a carrier concentration of 2.13·10$^{26}$ m$^{-3}$, which is very close to that of ITO thin film [10]. Cadmium oxide (CdO) is an exception, as it shows more than an order of magnitude higher carrier mobility compared to most of the TCO materials, and consequently an order of magnitude lower damping factor compared to other TCO materials [23, 24].

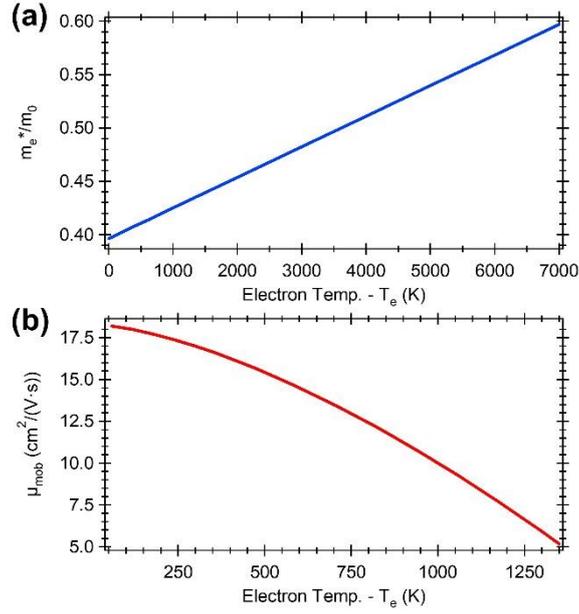

**Fig. 6**. (a) Effective mass and (b) carrier mobility as a function of electron temperature. Calculations were performed for the carrier concentration in ITO $N_c=1.5·10^{27}$ m$^{-3}$ and the corresponding effective mass $m^*=0.3964$.

A decrease in electron mobility with increasing electron temperature TCO is directly related to the increase in effective mass with light absorption due to the nonparabolic nature of the TCO materials. The total effective mass depends on the energy provided to the TCO material and is obtained by averaging the effective mass over the electron distribution in the conduction band:

$$\frac{1}{m^*} = \frac{1}{2\pi^2 m_0 N_c}\left(\frac{2m_0}{\hbar^2}\right)^{3/2} \int_0^\infty f_0 (E+CE^2)^{1/2}\, dE \tag{11}$$

where $f_0(E+CE^2)^{1/2}$ is the Fermi-Dirac distribution of electrons with the energy $E$ in the nonparabolic conduction band. For $C \rightarrow 0$, i.e., for a parabolic conduction band, the equation reduces to $m^*=m_0$. Based on it, the effective mass was calculated for a broad range of electron temperatures (Fig. 3a).

**Temperature coefficient of resistance (TCR)**

Under an absorption of light, the electron mobility decreases, which causes the electron conductivity to decrease. Thus, the change of mobility, and in a consequence, the effective mass can be monitored by a resistance measurements of TCO material under and absorption of light. *TCR* is a very valuable parameter that allows to evaluate the relative change of material resistance that is associated with a change in the material temperature and can be calculated from a formula:

$$TCR = \frac{\Delta R}{\Delta T}\frac{1}{R_0(T)} \tag{12}$$

where $\Delta R$ and $\Delta T$ are the changes in resistivity and temperature and the initial $R_0(T)$ is the resistivity at temperature $T$. It should be noted that fast thermal sensing requires materials with both high *TCR* and low thermal capacitance.

The resistance of TCO material can be calculated from a simple expression $R=(1/\sigma)(L/A)$, where $\sigma$ is the electrical conductivity, $L$ is the length and $A$ is the cross-sectional area of the TCO material. The electrical conductivity is expressed by $\sigma=N_c e\mu_{mob}$, where, $N_c$ is the carrier concentration, $e$ is the electron charge, and $\mu_{mob}$ is the carrier mobility. Assuming an intraband absorption of light, the electron mobility decreases as a result of the higher effective mass, which increases with the amount of light coupled to the TCO, while carrier concentration is kept at the same level. Thus, the *TCR* can be rewritten in a form:

$$TCR = -\frac{\Delta \mu}{\mu_{2,mob}} \frac{1}{\Delta T} \qquad (13)$$

where *Δμ_mob* is the change in the electron mobility and *μ_{2,mob}* is the electron mobility under a light coupled to the TCO. The change in the TCO resistance can be detected through the change of the photocurrent flowing through the TCO layer [41]:

$$I_{ph} = I_{off} - I_{on} = \frac{V_b}{R} - \frac{V_b}{R + \Delta R} = \frac{\Delta R}{R} \frac{U_b}{R + \Delta R} \qquad (14)$$

As observed, the photocurrent depends on the bias voltage $U_b$ and the relative change of the TCO resistance expressed by the ratio *ΔR/R*. Under a light illumination of the TCO, the effective mass increases causing a decrease in the electron mobility and, consequently, the electron conductivity decreases according to the equation *σ=neμ_mob*. From the knowledge of the optical power coupled to a photodetector and the photocurrent, the responsivity can be calculated. Calculations were performed for *L*=4 um long (distance between metal contacts) and *h*=10 nm thick TCO layer and under *P*=20 μW power coupled to the photodetector. The initial electron mobility, i.e., for a light off, was assumed to be *μ_mob*=18.3 cm²/(V·s), while the carrier concentration was assumed to be $N_c=1.5 \cdot 10^{27}$ m$^{-3}$.

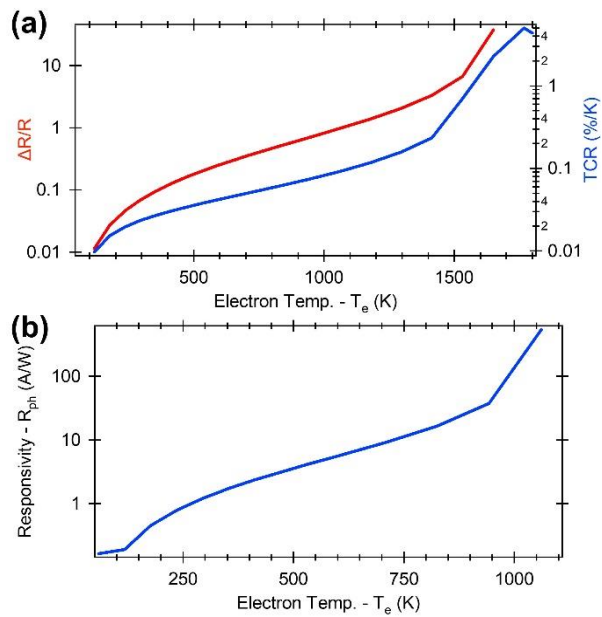

**Fig. 7.** (a) Resistance and temperature coefficient of resistance TCR change for different electron temperatures of ITO. (b) Calculated responsivity for an applied voltage of $U_b$=1 V as a function of electron temperature $T_e$.

As can be seen from Fig. 7, the relative resistance changes with increasing electron temperature, and hence the responsivity. However, it increases rapidly at the electron temperature around $T_e$=1000 K when the electron mobility under a light illumination approaches a half of the initial carrier mobility (for light off), defined here as *μ_mob*=18.3 cm²/(V·s). This is not surprising since the TCR grows exponentially with increasing electron temperature. Similar behavior as in Fig. 7a has been observed in VO_x-based bolometers where a steady increase in TCR was observed up to a certain temperature, beyond which a sharp increase was observed [42-43].

**Conclusion**

A new waveguide-integrated bolometric photodetector has been proposed that can operate in a broad wavelength range from NIR to MIR. This was possible owing to the broad absorption spectrum of the transparent conductive oxides, which were implemented for the first time as an active material in the phonic waveguide arrangement. It has been shown that under an absorption of light coupled to the photodetector by the TCO close to its ENZ point, the energy highly arises in TCO giving rise to the high electron temperature increases. Since the electron temperature scales with the electron effective mass and thus with the electron mobility, the change in optical power coupled to the photodetector can be monitored by the external electrical circuit. As a result, a theoretically predicted responsivity as high as tens of amperes per watt can be achieved. The compatibility of TCO with standard CMOS technology and the ability to build both modulators and photodetectors on the same material platform, it opens a new path in photonic integrated circuits.

**Acknowledgements**
J.G. thanks the "ENSEMBLE3 - Centre of Excellence for Nanophotonics, advanced materials and novel crystal growth-based technologies" project (GA No. MAB/2020/14) carried out within the International Research Agendas program of the Foundation for Polish Science co-financed by the European Union under the European Regional Development Fund and the European Union's Horizon 2020 research and innovation program Teaming for Excellence (Grant Agreement No. 857543) for support of this work.